%% file: ms.tex
\documentclass[conference]{IEEEtran}

\usepackage[letterpaper, left=1in, right=1in, bottom=1in, top=0.75in]{geometry}
\usepackage[pdftex]{graphicx}
\usepackage{amsmath}

\usepackage{tikz}
\usetikzlibrary{patterns}
\usetikzlibrary{shapes,arrows}
\usetikzlibrary{circuits.ee.IEC}
\usepackage{url}
\usepackage{amsthm, amssymb, cite}
\usepackage[utf8]{inputenc}

\usepackage{algorithm}% http://ctan.org/pkg/algorithms
\usepackage{algpseudocode}% http://ctan.org/pkg/algorithmicx
\usepackage[keeplastbox]{flushend}
\usepackage{caption}
\newtheorem{proposition}{Proposition}
\IEEEoverridecommandlockouts \IEEEpubid{\makebox[\columnwidth]{ 978-1-5386-3531-5/17/\$31.00~\copyright~2017 European Union \hfill} \hspace{\columnsep}\makebox[\columnwidth]{ }}
\renewcommand{\baselinestretch}{0.95}
\begin{document}
\title{Hybrid Beamforming with Spatial Modulation in Multi-user Massive MIMO mmWave Networks \thanks{The work of  Merve Yüzgeçcioglu has received partly funding from the European Union's Horizon 2020 research and innovation programme under the Marie Sklodowzka-Curie grant agreement No 641985}}
\author{\IEEEauthorblockN{Merve Yüzgeçcioglu and Eduard Jorswieck}
\IEEEauthorblockA{Communications Theory, Communications Laboratory\\
Dresden University of Technology,
D-01062 Dresden, Germany\\
Email: \{merve.yuzgeccioglu,\,eduard.jorswieck\}@tu-dresden.de}}
\maketitle
\begin{abstract}
The cost of radio frequency (RF) chains is the biggest drawback of massive MIMO millimeter wave  networks. By employing spatial modulation (SM), it is possible to implement lower number of RF chains than transmit antennas but still achieve high spectral efficiency. In this work, we propose a system model of the SM scheme together with hybrid beamforming at the transmitter and digital  combining at the receiver. In the proposed model, spatially-modulated bits are mapped onto indices of antenna arrays. It is shown that the proposed model achieves approximately 5dB gain over classical multi-user SM scheme with only 8 transmit antennas at each antenna array. This gain can be improved further by increasing the number of transmit antennas at each array without increasing the number of RF chains.
\end{abstract}
\IEEEpeerreviewmaketitle
\section{Introduction}
Recent technological developments also bring a challenge to wireless communications since the number of network devices increases drastically. Even though there are several works on achieving high data rates, the bandwidth shortage of today's commercial networks also limits the capacity of the channel. In order to overcome this bottleneck, millimeter wave (mmWave) spectrum became tempting due to the availability of large bandwidths. However, the signal in mmWave spectrum experiences a severe propagation loss and the resulting channel is poorly scattered. In spite of the unfavorable characteristics of the channel, beamforming  techniques can be employed at massive multiple-input multiple-output (mMIMO) networks to direct the beam with high array gain. Thanks to the small wave length at high frequencies, it is possible to pack a large amount of antennas in small areas and asymptotically achieve the capacity of the channel \cite{Marzetta2010}. However, packing high number of antennas comes with a price: energy consumption. It is not practical to implement dedicated RF chains for  each antenna at a transceiver due to the power and space restrictions. Nevertheless, by employing hybrid beamforming techniques, it is possible to benefit from the gain of a large amount of antennas while consuming less energy by using lower number of RF chains than antennas.

 There are several works on optimal precoding in hybrid beamforming \cite{Pi2012,Kim2013,Ayach2014,Alkhateeb2015,Sohrabi2015,Payami2016,Bogale2016}. In \cite{Ayach2014}, a method to construct a near-optimal hybrid beamforming structure is presented instead of an exhaustive search to maximize the spectral efficiency.  The work is extended to multi-user scenario in \cite{Alkhateeb2015} where users employ only analog combiner while the base station (BS) has a hybrid structure and constructs baseband precoder in order to mitigate the inter-user-interference. In \cite{Sohrabi2015}, together with a two-stage hybrid beamformer/combiner algorithm, the minimum number of RF chains to realize the performance with a fully-connected structure is given. Another approach is presented in \cite{Payami2016} that exploits the singular vectors of the channel to generate analog beamformer and combiner which provides a lower complexity. Furthermore, an asymptotic rate expression is given for the proposed scheme.

Another popular research topic recently is SM that enables transmitting additional bits to conventional modulated symbol without requiring extra power \cite{Mesleh2008,DiRenzo2014,Lee2015}. There have been extensive work on SM in different aspects for MIMO networks. Furthermore, implementation of SM at multi-user MIMO networks is studied at \cite{Narayanan2014} and a precoding method is proposed in order to cancel inter-user-interference. In this work, SM is employed at sub-groups which are dedicated for each user. Combination of SM and beamforming is introduced in \cite{Ishikawa2017}. In this work, analog beamforming is employed to a generalized SM (GSM) scheme for single-user scenario at Rician fading channels. A closer look into the transmitter design of SM together with analog beamforming is given in \cite{Lee2016}. Furthermore, the extension of SM with hybrid beamforming for a mmWave railway communication system is studied in \cite{Cui2016}. Here, the receive antenna arrays (AA) at the front and end of the train are assumed as virtual users served by two different data streams transmitted from the BS. Analog beamformer and combiner are employed at the transmitter and receiver, respectively. However, the beamformer and combiner are not designed specifically for multi-user spatial modulation system. Rather the design in \cite{Kim2013} is considered. As performance metric, an upper bound rate for AWGN channel with Gaussian input  is considered and maximum-likelihood detection is employed for decoding. However,  it is not clearly stated how the inter-user-interference is eliminated and data decoding is performed at each user.

In this paper, implementation of SM to a multi-user mmWave network is introduced. In this scheme, BS employs analog beamformer to direct the beam to the intended user and digital precoder to cancel inter-user-interference while users employ the digital combiners. The novelty of the paper is summarized as follows
\begin{itemize}
\item Analog beamforming is designed in order to maximize the achievable rate of the link. Two different codebooks are considered to choose analog beamformer vector such that one provides optimal beamformer while the other allows practical implementation. The quantization error is characterized between these two codebooks.
\item Digital precoder at the transmitter is designed in order to eliminate inter-user-interference. Digital combiner at the receiver is designed to successfully reconstruct transmitted symbol when it is matched with the correct analog beamformer.
\item A low-complexity maximum-likelihood detector is introduced.
\item Analytical expression of the achievable rate of the system is derived and the tightness of the expression is shown.
\end{itemize}
The rest of the paper is organized as follows. In Sec.~\ref{sec:SystemModel}, the system model of the proposed scheme is described. In Sec.~\ref{sec:Design}, design of the analog beamformer and digital precoder for transmitter side and the digital combiner for receiver side is explained in detail, and the quantization error between two codebooks is characterized. In Sec.~\ref{sec:Rate}, analytical expression of achievable rate for the proposed scheme is derived. The performance is analyzed numerically in Sec.~\ref{sec:NumericalResults} and finally the paper is concluded in Sec.~\ref{sec:Conclusion}.
\section{System Model}
\label{sec:SystemModel}
\begin{figure*}
\centering
\input{HBFSM_BlockDiagram.tikz}
\caption{Block diagram of hybrid beamforming with spatial modulation in multi-user downlink transmission}
\label{fig:BlockDiagram}
\end{figure*}
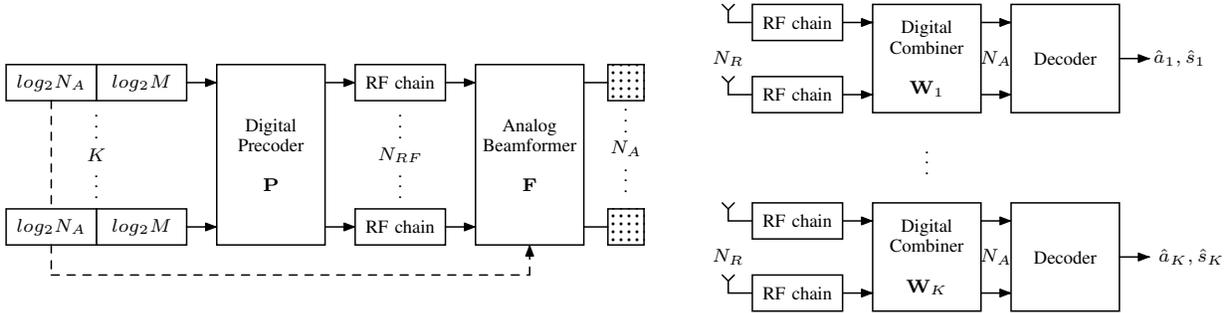
The block diagram of the system is shown in Fig.~\ref{fig:BlockDiagram}. There, the BS is equipped with $N_A$ uniform linear arrays (ULA), each with $N_T$ transmit antennas connected to $N_{RF}$ RF chains. We assume without loss of generality that $N_T$ is fixed for all AAs and there are no common transmit antennas between arrays. The BS serves $K$ users with $N_R$ receive antennas and it is assumed that $N_{RF} \geq K$. The number of AAs may vary independently from the number of RF chains at the transmitter.

Data is transmitted to mobile stations (MS) in two parts: first part is modulated according to spatial modulation principles by choosing one of the AAs which allows to transmit $\log_2 N_A$ bits and the second part is modulated according to a conventional $M$-ary modulation scheme which allows to transmit $\log_2 M$ bits.  The index of the active AA conveys information without consuming additional power.  According to incoming bits for each bit stream, it is possible that one AA is used for more than one user at a channel use. This is only possible by the help of analog beamformer and allows to exploit SM for multi-user scenario without any limitation. After determining the AA indices for every user, the transmitter constructs the digital precoder and the analog beamformer. The received signal at $i$-th user can be expressed as follows
%\vspace{-.2cm}
\begin{equation}
\mathbf{r}_i = \sqrt{\rho}\sum_{j=1}^K \mathbf{H}_{a_j,i}\mathbf{f}_{a_j,j}\mathbf{p}_{a_j,j}\mathbf{s} + \mathbf{n}_i.
\end{equation}
Herein, $\mathbf{H}_{a_j,i}\in\mathbb{C}^{N_R\times N_T}$ is the $L$-path channel between the $a_j$-th AA and $i$-th user where $a_j$ is the selected AA index to transmit data to $j$-th user. The channel follows a geometry-based model shown below \cite{Saleh1987}
%\vspace{-.2cm}
\begin{equation}
\mathbf{H}_{a_j,i} = \sqrt{\frac{N_TN_R}{L}}\sum_{l=1}^{L}\alpha_{a_j,i_l}\mathbf{a}_{R_{a_j,i}}(\theta_l)\mathbf{a}_{T_{a_j,i}}^H(\phi_l),
\label{eq:Channel}
\end{equation}
Here, $\mathbf{a}_{R_{a_j,i}}(\theta_l)\!=\!\sqrt{\frac{1}{N_R}}\![1\!,\!\mathrm{e}^{j\pi\sin\theta_l}\!,\!\dots\!,\!\mathrm{e}^{j\pi(N_R-1)\sin\theta_l}\!]^T$ and $\mathbf{a}_{T_{a_j,i}}(\phi_l)\!=\!\sqrt{\frac{1}{N_T}}\![1\!,\!\mathrm{e}^{j\pi\sin\phi_l}\!,\!\dots\!,\!\mathrm{e}^{j\pi(N_T-1)\sin\phi_l}\!]^T$ are the receive and transmit AA responses of the $l$-th path where $l = 1,\dots,L$.  $\phi_l$ and $\theta_l$ are the angle of departure (AoD) and angle of arrival (AoA) of the path and drawn from the uniform distribution $\mathcal{U}(0,2\pi]$. Finally, $\alpha_{a_j,i_l}$, $\mathcal{CN}(0,1)$, represents the channel gain and the path loss of the path. Throughout the paper, $(.)^T$ and $(.)^H$ denote transpose and hermitian operators, respectively. It is assumed that the antennas are placed on the array with half-wavelength separation. $\mathbf{p}_{a_j,j} \in \mathbb{C}^{1\times K}$ and $\mathbf{f}_{a_j,j} \in \mathbb{C}^{N_T\times 1}$ are the digital precoder and analog beamformer vectors for $j$-th user when $a_j$-th AA array is chosen for transmission. $\rho$ represents the total transmit power and with equal power allocation, the transmit power of $i$-th user is $\rho_i = \rho/K$. The selection procedure for precoder and beamformer is explained in detail in Sec. \ref{sec:Design}. $\mathbf{s} \in \mathbb{C}^{K\times 1}$ is the data vector contains symbols for K users and finally  $\mathbf{n}_i \in \mathbb{C}^{N_R\times 1}$ is the noise vector with elements distributed as $\mathcal{CN}(0,\sigma^2)$.
\subsection{Receiver Structure}
\label{sec:ReceiverStructure}
The receiver structure of the users is shown  in Fig.~\ref{fig:BlockDiagram}. In this system model, MS has $N_R$ receive antennas connected to the same number of RF chains. The received signal is passed on to the combiner with the help of RF chains and digital combining is performed to successfully reconstruct the transmitted symbol. In order to decode the received symbol, a low complexity maximum-likelihood (ML) detector is employed to jointly estimate the AA index and $M$-ary modulated symbol
%\vspace{-.2cm}
\begin{equation}
[\hat{a}_i, \hat{s}_i] = \text{arg}\displaystyle\min_{a,m}{\left|\frac{1}{\beta}\mathbf{w}_{a,i}^H\mathbf{r}_i-s_m\right|^2},
\end{equation}
where $\mathbf{w}_{a,i} \in \mathbb{C}^{N_R \times 1}$ is the digital combiner from $a$-th AA to $i$-th user, $a = 1,\dots,A$, $\beta$ is the normalization coefficient and $s_m$ is the $m$-th symbol from $M$-ary constellation diagram. This detector structure allows the user to decode without having $\mathbf{H}_{a_j,i}$, $\mathbf{f}_{a_j,j}$ and $\mathbf{p}_{a_j,j}$ knowledge but only $\mathbf{w}_{a,i}$ vectors and $\beta$. The construction of $\mathbf{w}_{a,i}$ is also explained in detail in Sec. \ref{sec:Design}.
\section{Precoder, Beamformer and Combiner Design}
\label{sec:Design}
We consider two different approaches to design the analog beamformer vector. The first method is choosing the vector among the transmit AA response vectors $\mathbf{a}_T(\phi)$. The second method is using a predefined beamsteering codebook with quantized angles between $(0,2\pi]$.

Let us define the codebook $\mathcal{F}=\{\mathbf{f}_n \in \mathbb{C}^{N_T\times 1}:\mathbf{f}_n^H\mathbf{f}_n = 1,\, n = 1,\dots,N\}$  and name the two different codebooks as $\mathcal{F}_A$ for transmit AA response vectors collection and $\mathcal{F}_B$ for the beamsteering codebook. For both cases, the individual codewords are constructed as $\mathbf{f}=\sqrt{\frac{1}{N_T}}[1,\exp{(j\pi\phi)},\dots,\exp{(j\pi(N_T-1)\phi)}]^T$ where $\phi$ is the AoD of each path for codebook $\mathcal{F}_A$ and quantized angle $\frac{2\pi n}{N}$ where $n = 0,\dots,N-1$ for codebook $\mathcal{F}_B$.

Furthermore, the analog beamformer $\mathbf{f}$ is chosen in order to maximize the signal-to-noise ratio (SNR) of the link between the selected AA and the intended user. For each user, the BS computes the RF beamformer $\mathbf{f}_{1,i},\dots,\mathbf{f}_{N_A,i}$ as follows
%\vspace{-.2cm}
\begin{equation}\label{eq:Optimization}
\mathbf{f}_{a,i} = \text{arg}\displaystyle\max_{\forall \mathbf{f}_n \in \mathcal{F}}{||\mathbf{H}_{a,i}\mathbf{f}_n||^2},
\end{equation}
where $\mathbf{f}_{a,i}$ is the optimum RF beamformer from the $a$-th AA to $i$-th user. The resulting analog beamforming matrix for all users is $\mathbf{F}=[\mathbf{f}_{a_1,1},\dots,\mathbf{f}_{a_K,K}]$.

After determining the optimum RF beamformers for each AA-user pair, the BS constructs the matrix $\mathbf{H}_i = [\mathbf{H}_{1,i}\mathbf{f}_{1,i},\dots,\mathbf{H}_{N_A,i}\mathbf{f}_{N_A,i}]$. $\mathbf{H}_i$ is used to calculate the digital combiner for each user to successfully  reconstruct the transmitted symbol
%\vspace{-.2cm}
\begin{equation}
\mathbf{W}_i =
 \begin{bmatrix}\mathbf{w}_{1,i}\dots\mathbf{w}_{N_A,i}\end{bmatrix} = (\mathbf{H}_i^\dag)^H,
\end{equation}
where $(.)^\dag$ denotes the pseudo inverse of a matrix. With this design, it is possible to feedback only $\mathbf{H}_i$ matrices to the users instead of having a training period of the channel matrices for each AA-user pair. In this case, it is sufficient for users to have $\mathbf{H}_i$ with dimension $N_A\times N_R$ instead of full channel matrices $\mathbf{H}_{1,i},\dots,\mathbf{H}_{N_A,i}$ with dimension $N_AN_T\times N_R$.

After determining the optimum RF beamformer and receive combiner, the transmitter generates the effective channel in Eq.~[\ref{eq:EffChan}] order to calculate the digital precoder to eliminate the inter-user-interference
%\vspace{-.2cm}
\begin{equation}\label{eq:EffChan}
\mathbf{H}_{eff}\!=\!\begin{bmatrix}\!
\!\mathbf{w}_{a_1,1}^H\mathbf{H}_{a_1,1}\mathbf{f}_{a_1,1}\!&\!\dots\!&\!\mathbf{w}_{a_1,1}^H\mathbf{H}_{a_K,1}\mathbf{f}_{a_K,K}\\
\!\vdots\!&\!\ddots\!&\!\vdots\\
\!\mathbf{w}_{a_K,K}^H\mathbf{H}_{a_1,K}\mathbf{f}_{a_1,1}\!&\!\dots\!&\!\mathbf{w}_{a_K,K}^H\mathbf{H}_{a_K,K}\mathbf{f}_{a_K,K}\!
\end{bmatrix}.
\end{equation}
Finally the precoding vector for each user is determined as $\mathbf{P}=[\mathbf{p}_{a_1,1}^T\dots\mathbf{p}_{a_K,K}^T]^T = \beta\mathbf{H}_{eff}^\dag$,where $ \mathbf{p}_{a_k,k}\in\mathbb{C}^{1\times K}$ is the precoder vector of $k$-th user when $a_k$-th AA is chosen by spatially-modulated bits.  The coefficient $\beta$ satisfies the average power constraint which is calculated as
%\vspace{-.2cm}
\begin{equation}
\beta = \mathbb{E}\left\{\sqrt{\frac{K}{tr(\mathbf{H}_{eff}^\dag(\mathbf{H}_{eff}^\dag)^H)}}\right\},
\end{equation}
where the expectation is over channel realizations. Since the complex channel coefficients are the key parameters for spatially-modulated bits, instantaneous power constraint is not applicable for such systems \cite{Narayanan2014}.
\subsection{Characterization of the Quantization Error}
\label{sec:Error}
For large scale AAs, the right and left singular vectors of the channel are the transmit and receive AA response vectors, respectively \cite{Ayach2012}. Let us define the optimal analog beamformer for the channel given in Eq.~[\ref{eq:Channel}] in accordance with this statement as follows
%\vspace{-.2cm}
\begin{equation}
\mathbf{f}_A = \text{arg}\displaystyle\max_{\forall \mathbf{f}_{A,n} \in \mathcal{F}_A}{||\mathbf{H}\mathbf{f}_{A,n}||^2},
\end{equation}
where $N=L$ is the number of paths of the channel and the total number of elements in the codebook. $\mathbf{H}\in \mathbb{C}^{N_R\times N_T}$ is the channel matrix, $\mathbf{f}_{A,n}\in \mathbb{C}^{N_T\times 1}$ is the $n$-th element of the codebook $\mathcal{F}_A$. In order to achieve the performance of the optimal beamformer, infinite resolution phase shifters should be implemented at the system. Since this is not possible for practical systems, beamsteering codebooks with quantized angle are commonly considered. However, selecting the analog beamformer among beamsteering codebooks leads to a quantization error.
\begin{proposition}\label{prop:QE}
	Assume that $\mathbf{f}_A$ and $\mathbf{f}_B$ are the solution to Eq.~[\ref{eq:Optimization}] for codebooks $\mathcal{F}_A$ and $\mathcal{F}_B$, respectively. Then, the quantization error can be characterized as follows
	%\vspace{-.2cm}
	\begin{equation}
	d_{c,min} \leq \max_{\forall \mathbf{f}_A \in G(N_T,1)}{d_c^2(\mathbf{f}_A,\mathbf{f}_B)}\leq d_{c,max},
	\label{eq:error}
	\end{equation}
	where $d_{c,min} = 1-tr\{\mathbf{f}_A\mathbf{f}_A^H\mathbf{f}_{B,n}\mathbf{f}_{B,n}^H\}$, $d_{c,max} = c^{-\frac{1}{N_T-1}}2^{-\frac{B}{N_T-1}}$ and $c$ is the coefficient of the metric ball volume defined in \cite{Krishnamachari2013}.	
\end{proposition}
\begin{IEEEproof}
	Let us assume the codebook $\mathcal{F}_B$ with $B$ bits resolution phase shifters where codebook size is $N=2^B$. Eq.~[\ref{eq:Optimization}] can be reformulated as follows
	%\vspace{-.2cm}
	\begin{equation}
	\begin{split}
	\mathbf{f}_B &\triangleq \text{arg}\displaystyle\min_{\forall \mathbf{f}_{B,n} \in \mathcal{F}_B}{d_c^2(\mathbf{f}_A,\mathbf{f}_{B,n})} \\
	&= \text{arg}\displaystyle\min_{\forall \mathbf{f}_{B,n} \in \mathcal{F}_B}{1-tr\{\mathbf{f}_A\mathbf{f}_A^H\mathbf{f}_{B,n}\mathbf{f}_{B,n}^H\}},
	\end{split}
	\label{eq:Distance}
	\end{equation}
	where $d_c(.,.)$ is the chordal distance. Eq.~[\ref{eq:Distance}] is an example of Grassmannian quantization on the Grassmann manifold $G(N_T,1)$. The codebook $\mathcal{F}_B$ can be considered as a Grassmannian subspace sphere-packing codebook since the elements of  $\mathcal{F}_B$ are distributed over a sphere with radius $||\mathbf{f}_{B,n}^H\mathbf{f}_{B,n}||^2=1$, $\forall n=1,\dots,N$. Let us define the distance between the optimal beamformer and the selected beamformer from the quantized codebook as $d_c(\mathbf{f}_A,\mathbf{f}_B)$. By using the results in \cite{Dai2008,Krishnamachari2013,Cao2016} we reach the Eq.~[\ref{eq:error}].
\end{IEEEproof}
 It is clearly seen from Eq.~[\ref{eq:error}] that the quantization error vanishes asymptotically such as $\lim_{B\rightarrow \infty} c^{-\frac{1}{N_T-1}}2^{-\frac{B}{N_T-1}} = 0$.
\section{Achievable Rate}
\label{sec:Rate}
\begin{figure}
\centering
\includegraphics[scale=0.4]{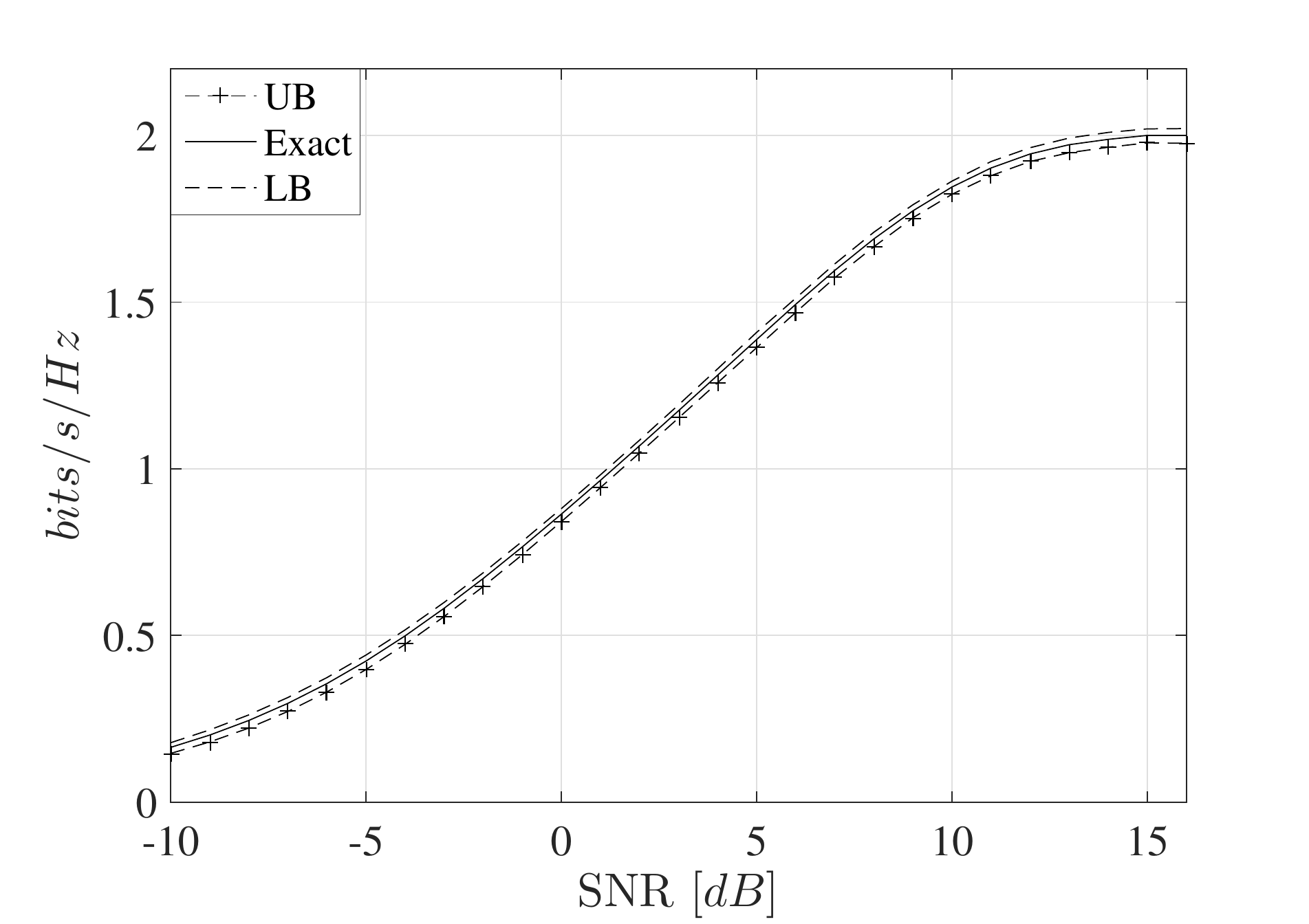}
\caption{Rate approximation of a user for $N_A = 2$, $N_T = 8, N_R = 1$ with BPSK modulation}
\label{fig:Rate}
\end{figure}
The received signal at $i$-th user after receive combining is
%\vspace{-.2cm}
\begin{equation}
\begin{split}
y_i=\mathbf{w}_{a_i,i}^H\sqrt{\rho}\sum_{j=1}^K\mathbf{H}_{a_j,i}\mathbf{f}_{a_j,j}\mathbf{p}_{a_j,j}\mathbf{s}+\mathbf{w}_{a_i,i}^H\mathbf{n}_i.
\end{split}
\end{equation}
Since the interference from the other users is eliminated by employing transmit precoder, the resulting signal at the receiver can be reformulated as follows
%\vspace{-.2cm}
\begin{equation}
y_i = \sqrt{\rho_i}\mathbf{w}_{a_i,i}^H\mathbf{H}_{a_i,i}\mathbf{f}_{a_i,j}s_i + \mathbf{w}_{a_i,i}^H\mathbf{n}_i.
\end{equation}
Information theoretical methods can be used to find the achievable rate of the proposed scheme. The mutual information between transmitted and received symbols is written as
%\vspace{-.2cm}
\begin{equation}
I(y_i;x_i) = h(y_i) - h(z_i).
\end{equation}
Therein, $x_i$ is considered as the symbol that contains the AA index and $M$-ary modulated symbol to ease the understanding. $z_i$ is the noise term after receive combining which follows the same distribution with $\mathbf{n}_i$.
\begin{proposition}\label{prop:Rate}
	The received signal $y_i$ follows Gaussian mixture distribution. There is no closed-form solution for the entropy of a Gaussian mixture but using a tight approximation the achievable rate can be bounded as
%\vspace{-.2cm}
	\begin{equation}\label{eq:rate}
	h_l(y_i) - h(z_i) \leq R_i \leq h_u(y_i) - h(z_i),
	\end{equation}
	where $h_l(y_i)=(\gamma+\alpha_{N,N'}) \log_2e + \log_2\sigma$, $h_u(y_i)=(\gamma+\beta_{N,N'}) \log_2e + \log_2\sigma$ and $h(z_i)=\log_2(\pi e \sigma^2)$.
\end{proposition}
\begin{IEEEproof}
	We begin the proof by deriving the probability density function of the received signal that can be written as follows
	%\vspace{-.2cm}
	\begin{equation}\label{eq:chain}
	f_{Y_i}\!(\! y_i\!)\!=\!\sum\limits_{m=1}^M\!\sum\limits_{j=1}^{N_A}\!f_{Y_i}\!(\! y_i|a\!=\! j\!,\! s\!=\!s_m\!)\! P\!(a\!=\! j\!,\! s\!=\! s_m\!)\!.
	\end{equation}
	Since the selection of the AA indices and $M$-ary symbols are independent events, joint probability in Eq.~[\ref{eq:chain}] can be separated as $P(a = j, s = s_m) = P(a = j)P(s = s_m) = \frac{1}{N_A}\frac{1}{M}$. Furthermore, it is clearly seen that $f_{Y_i}(y_i|a = j, s = s_m)$ follows complex Gaussian distribution with different mean values $\sqrt{\rho_i}\mathbf{w}_{j,i}^H\mathbf{H}_{j,i}\mathbf{f}_{j,i}s_m$ according to the $j$ and $m$ indices. The resulting probability density function follows Gaussian mixture distribution
	%\vspace{-.2cm}
	\begin{equation}
	\begin{split}
	f_{Y_i}(y_i) =&\frac{1}{MN_A}\sum\limits_{m=1}^M\sum\limits_{j=1}^{N_A}\\
	 &\frac{1}{\pi}\!\exp\!\left\{\!-\!\frac{|y_i-\sqrt{\rho_i}\mathbf{w}_{j,i}^H\mathbf{H}_{j,i}\mathbf{f}_{j,i}s_m|^2}{\sigma^2}\!\right\}\!,
	\end{split}
	\end{equation}
	where $\mathbf{H}_{j,i}$, $\mathbf{f}_{j,i}$ and $\mathbf{w}_{j,i}$ are the channel matrix, analog beamformer and receive combiner for $j,i$ AA-user pair, respectively. Herein, $s_m$ denotes the $m$-th symbol from $M$-ary constellation diagram and $\rho_i$ is the transmit power for $i$-th user. By using the tight approximation derived for Gaussian mixtures in \cite{Moshksar2016}, entropy of $y_i$ can be bounded as $(\gamma+\alpha_{N,N'}) \log_2e + \log_2\sigma \leq h(y_i) \leq (\gamma+\beta_{N,N'}) \log_2e  + \log_2\sigma$. Due to space constraints the detailed explanation on the parameters $\gamma$, $\alpha_{N,N'}$ and $\beta_{N,N'}$ are not given here and can be found in \cite{Moshksar2016}.
\end{IEEEproof}

The tightness of the introduced approximation is shown in Fig.~\ref{fig:Rate}. In this figure, the lower (LB) and upper (UB) bounds are calculated by using the approximation and the exact rate is found by calculating the exact integral for entropy $h(y) = -\int_{-\infty}^{\infty} f_Y(y)\log f_Y(y)dy$.
\section{Numerical Results}
\label{sec:NumericalResults}
In this section, the performance of the hybrid beamforming with spatial modulation (HBF-SM) is analyzed numerically. Firstly, the effect of the quantization error on the uncoded bit error rate (BER) performance is investigated for different codebook sizes. Then, it is compared with the conventional SM scheme for multi-user scenario  \cite{Narayanan2014}.
\begin{figure}
\centering
\includegraphics[scale=0.4]{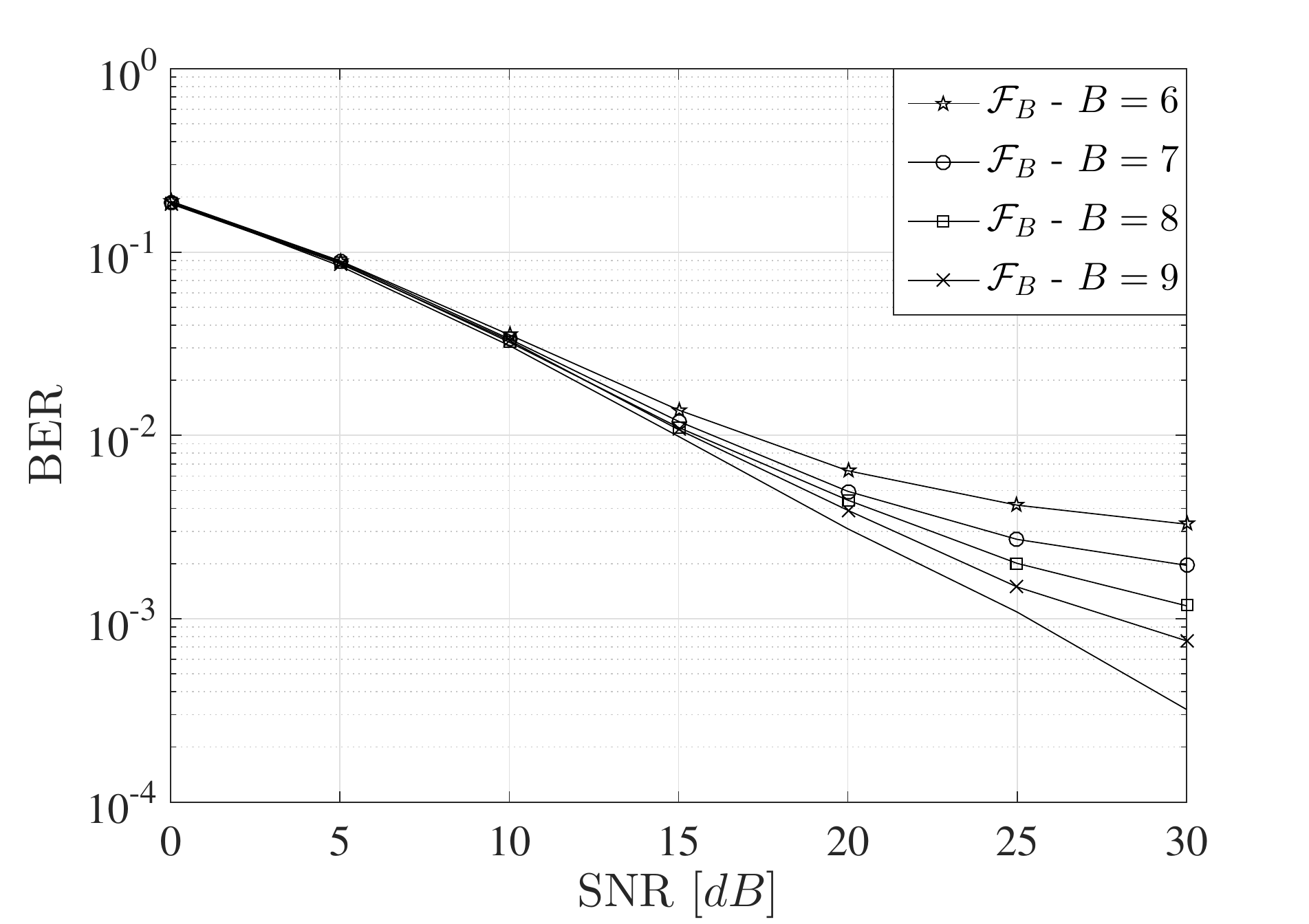}
\caption{BER performance for different codebook sizes ($K = 2, N_A = 4, N_T = 8, N_R = 1, 4-QAM$)}
\label{fig:HBFSM_different_CS}
\end{figure}

In Fig.~\ref{fig:HBFSM_different_CS}, the BER performance of the system is shown for the cases that the analog beamformer vector is chosen among the transmit AA response vectors ($\mathcal{F}_A$) and the beamsteering codebook ($\mathcal{F}_B$). It is seen that the quantization error of the codebook $\mathcal{F}_B$ leads to an error floor for the low codebook sizes. Since the digital combiner at the receiver side is generated by using the analog beamforming vector also, when  the beamforming vector is erroneous, it cumulatively increases the resulting error probability. As it is expected, when the analog beamformer vector is chosen from codebook $\mathcal{F}_A$, the performance is better and there is no error floor in the system. Although the codebook $\mathcal{F}_A$ provides the optimal beamformer, there is no phase shifter  with infinite resolution. Hence, the codebook $\mathcal{F}_B$ with $B$-bit resolution would be used in practical systems. When the phase shifters are changed from $6$-bits to $9$-bits the performance is increased significantly. As it is shown in Sec.~\ref{sec:Error} the performance of the beamsteering codebook will eventually converge to the optimal beamformer.
\begin{figure}
\centering
\includegraphics[scale=0.4]{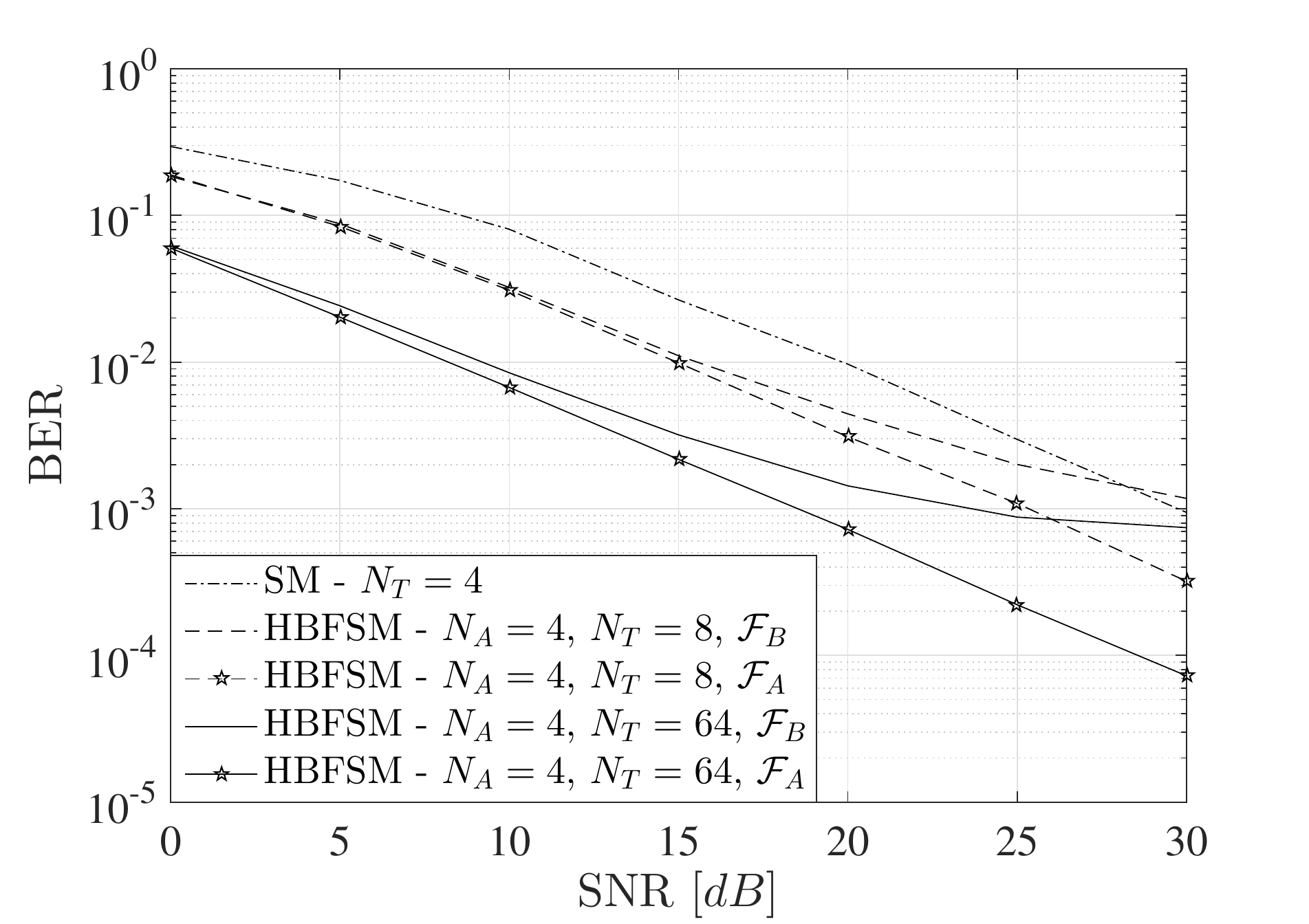}
\caption{Comparison of classical SM and HBF-SM ($K = 2, N_R = 1, B = 8, 4-QAM$)}
\label{fig:MUSMvsHBFSM4}
\end{figure}

In Fig.~\ref{fig:MUSMvsHBFSM4}, the BER performance of the HBF-SM is compared with the conventional SM scheme in multi-user scenario. Average power constraint is employed for both schemes. The considered SM scheme has $N_T = 4$ transmit antennas. For HBF-SM scheme, $N_A = 4$ antenna arrays are considered with $N_T = 8$ and $N_T = 64$ transmit antennas at each array, respectively. In SM scheme, total $\log_2 N_T = 2$ bits are transmitted at a channel use by the selected transmit antenna index. In HBFM-SM scheme, also $\log_2 N_A = 2$ bits are transmitted by the selected AA index. It is seen from the simulation results, although HBF-SM has an error floor when codebook $\mathcal{F}_B$ is used, the error performance is better than the classical SM scheme up to high SNR regime. When the optimal codebook is used, HBF-SM outperforms classical SM regardless the SNR value. It is possible to have $5dB$ gain over classical SM scheme with only $8$ transmit antennas at an AA. If we consider  $64$ transmit antennas at each array, the gain will increase up to $12dB$. Although there is an intersection point between HBFM-SM and classical SM, by increasing the codebook size and transmit antenna numbers at an array, the performance of HBF-SM can be further increased. Note that, each AA is served by a single RF chain. Hence, increasing the number of transmit antennas at an array does not come with a high cost while significantly increasing the performance. As it is seen from the simulation results, the number of antennas at an array is one of the key parameters that affects the performance. The larger the number of transmit antennas the better the beam can be directed to intended direction.
\section{Conclusion}
\label{sec:Conclusion}
In this work, we have introduced a system model of HBF-SM scheme. Novel design of the beamformer and digital combiners for the proposed model have been derived. The effect of the codebook selection for analog beamformer  on performance has been shown and the quantization error has been studied. Additionally, a tight approximation for the rate has been introduced. Uncoded BER performance of the system has been shown for different parameters and also compared with a classical SM implementation for multi-user networks. Simulation results showed that even with low number of transmit antennas at the AAs, performance of classical SM is outperformed.
For future work, the receiver design can be improved by employing hybrid beamforming that also allows reducing the number of RF chains. Since the most important objective of the SM schemes is providing higher spectral efficiency while consuming lower energy compared to conventional schemes, optimizing the energy efficiency of the system is an important direction for future research.
%\section*{Appendix}
\renewcommand{\baselinestretch}{0.95}
% $ biblatex auxiliary file $
% $ biblatex bbl format version 2.6 $
% Do not modify the above lines!
%
% This is an auxiliary file used by the 'biblatex' package.
% This file may safely be deleted. It will be recreated by
% biber as required.
%
\bibliographystyle{IEEEtran}
\bibliography{article}
%\printbibliography

\flushend
\end{document}

%% file: HBFSM_BlockDiagram.tikz
\usetikzlibrary{circuits.ee.IEC}
\usetikzlibrary{arrows,decorations.markings}

%%%%%%%%%%%%%%%%%%%%%%%%%%%%%%%%%%%%%%%%%%%%%%%%%%%%%%%%%%%%%%%%%%%%%%%%%%%%%%%%%%%%%%%%%

\begin{tikzpicture}[circuit ee IEC, scale=0.8, line width=0.02cm, line cap=round, font=\scriptsize]

	% Bit Stream
	
	\filldraw[fill=white, draw=black] (-9.5,1.5) rectangle (-8,0.9);
	\node[align=center] at (-8.75,1.2) {$log_2M$};
	\filldraw[fill=white, draw=black] (-11,1.5) rectangle (-9.5,0.9);
	\node[align=center] at (-10.25,1.2) {$log_2N_A$};
	
	\node[align=center] at (-9.5,0.6) {$\vdots$};
	\node[align=center] at (-9.5,0) {$K$};
	\node[align=center] at (-9.5,-0.4) {$\vdots$};
	
	\filldraw[fill=white, draw=black] (-9.5,-1.5) rectangle (-8,-0.9);
	\node[align=center] at (-8.75,-1.2) {$log_2M$};
		\filldraw[fill=white, draw=black] (-11,-1.5) rectangle (-9.5,-0.9);
	\node[align=center] at (-10.25,-1.2) {$log_2N_A$};
	
	% SM part to ABF
	\draw[dashed] (-10.25,0.9) -- (-10.25,-0.9);
	\draw[dashed] (-10.25,-1.5) -- (-10.25,-2);
	\draw[dashed] (-10.25,-2) -- (-2.3,-2);
	\draw[dashed] (-2.3,-2) -- (-2.3,-1.5);
	\node[current direction={rotate = 90}] at (-2.3,-1.6) {};
	
		% Wiring bitstream to Digital Precoder
	\draw (-8,1.2) -- (-7.5,1.2);
	\node[current direction={rotate = 0}] at (-7.6,1.2) {};
	
	\draw (-8,-1.2) -- (-7.5,-1.2);
	\node[current direction={rotate = 0}] at (-7.6,-1.2) {};
	
	% Digital Precoder block
	\draw (-7.5,1.5) rectangle (-5.7,-1.5);
	\node[align=center] at (-6.6,0) {Digital\\Precoder\\ \\$\mathbf{P}$};
	
	% Wiring Digital Precoder to Tx RF chains
	\draw (-5.7,1.2) -- (-5.2,1.2);
	\node[current direction={rotate = 0}] at (-5.3,1.2) {};
	
	\draw (-5.7,-1.2) -- (-5.2,-1.2);
	\node[current direction={rotate = 0}] at (-5.3,-1.2) {};
	
		% Tx RF chains
	
	\filldraw[fill=white, draw=black] (-5.2,1.5) rectangle (-3.7,0.9);
	\node[align=center] at (-4.45,1.2) {RF chain};
	
	\node[align=center] at (-4.45,0.6) {$\vdots$};
	\node[align=center] at (-4.45,0) {$N_{RF}$};
	\node[align=center] at (-4.45,-0.4) {$\vdots$};
	
	\filldraw[fill=white, draw=black] (-5.2,-1.5) rectangle (-3.7,-0.9);
	\node[align=center] at (-4.45,-1.2) {RF chain};
	
	% Wiring Tx RF chains to Analog Beamformer
	\draw (-3.7,1.2) -- (-3.2,1.2);
	\node[current direction={rotate = 0}] at (-3.3,1.2) {};
	
	\draw (-3.7,-1.2) -- (-3.2,-1.2);
	\node[current direction={rotate = 0}] at (-3.3,-1.2) {};
	
	% Analog Beamformer Block
	\filldraw[fill=white, draw=black] (-3.2,1.5) rectangle (-1.4,-1.5);
	\node[align=center] at (-2.3,0) {Analog\\Beamformer\\ \\$\mathbf{F}$};
	
	% Antenna Arrays
	\draw (-1.4,1.2) -- (-1,1.2);
	\filldraw[fill=white, draw=black, pattern=dots, pattern color=black] (-1,1.5) rectangle (-0.4,0.9);
	
	\node[align=center] at (-0.7,0.7) {$\vdots$};
	\node[align=center] at (-0.7,0.1) {$N_A$};
	\node[align=center] at (-0.7,-0.3) {$\vdots$};
	
	\draw (-1.4,-1.2) -- (-1,-1.2) ;
	\filldraw[fill=white, draw=black, pattern=dots, pattern color=black] (-1,-1.5) rectangle (-0.4,-0.9);

	% Radio Channel
	%\filldraw[fill=black!8!white, draw=black!8!white] (-5.3,1.7) rectangle (5.3,-2.2);
	%\node[align=center] at (0,-1.95) {effective channel $\tilde{\mathbf{H}}$};
	
	%\filldraw[fill=black!16!white, draw=black!16!white] (-1.2,1.6) rectangle (1.2,-1.6);
	%\node[align=center] at (0,0) {radio\\channel\\ \\ \\$\mathbf{H}$};
	
	%%%%%%%%%%%%%%%%%
	% 1st user
		% Rx antennas	
	\draw (1.4,2.2) -- (1,2.2) -- (1,2.4) -- (1.1,2.5);
	\draw (1,2.4) -- (0.9,2.5);
	
	%\node[align=center] at (1,0.7) {$\vdots$};
	\node[align=center] at (1,1.6) {$N_R$};
	%\node[align=center] at (1,-0.3) {$\vdots$};
	
	\draw (1.4,1) -- (1,1) -- (1,1.2) -- (1.1,1.3);
	\draw (1,1.2) -- (0.9,1.3);
	
		% Rx RF chains 	
	\filldraw[fill=white, draw=black] (2.9,2.5) rectangle (1.4,1.9);
	\node[align=center] at (2.15,2.2) {RF chain};
	
	\filldraw[fill=white, draw=black] (2.9,0.7) rectangle (1.4,1.3);
	\node[align=center] at (2.15,1) {RF chain};
	
	% Wiring Rx RF chains to Digital combiner
	\draw (2.9,2.2) -- (3.4,2.2);
	\node[current direction={rotate = 0}] at (3.3,2.2) {};
	
	\draw (2.9,1) -- (3.4,1);
	\node[current direction={rotate = 0}] at (3.3,1) {};
	
	% Digital Combiner block	
	\draw (3.4,2.5) rectangle (5.2,0.7);
	\node[align=center] at (4.3,1.6) {Digital\\Combiner\\ \\$\mathbf{W}_1$};
	
	% Wiring Digital Combiner to Decoder
	
	\draw (5.7,2.2) -- (5.2,2.2);
	\node[current direction={rotate = 0}] at (5.6,2.2) {};
	
	%\node[align=center] at (5.45,0.7) {$\vdots$};
	\node[align=center] at (5.45,1.6) {$N_A$};
	%\node[align=center] at (5.45,-0.5) {$\vdots$};
	
	\draw (5.7,1) -- (5.2,1);
	\node[current direction={rotate = 0}] at (5.6,1) {};

	% Decoder Block
	\filldraw[fill=white, draw=black] (5.7,2.5) rectangle (7.5,0.7);
	\node[align=center] at (6.6,1.6) {Decoder};

	% Decoder output
	\draw (8,1.6) -- (7.5,1.6);
	\node[current direction={rotate = 0}] at (7.9,1.6) {};		
	\node[align=center] at (8.5,1.6){$\hat{a}_1, \hat{s}_1$};

	\node[align=center] at (4.3,0) {$\vdots$};

	%%%%%%%%%%%%%%%%%%%%%%%%%%%%
	% K-th user
		% Rx antennas	
	\draw (1.4,-1.1) -- (1,-1.1) -- (1,-0.9) -- (1.1,-0.8);
	\draw (1,-0.9) -- (0.9,-0.8);
	
	%\node[align=center] at (1,0.7) {$\vdots$};
	\node[align=center] at (1,-1.7) {$N_R$};
	%\node[align=center] at (1,-0.3) {$\vdots$};
	
	\draw (1.4,-2.3) -- (1,-2.3) -- (1,-2.1) -- (1.1,-2.0);
	\draw (1,-2.1) -- (0.9,-2.0);
	
		% Rx RF chains 	
	\filldraw[fill=white, draw=black] (2.9,-0.8) rectangle (1.4,-1.4);
	\node[align=center] at (2.15,-1.1) {RF chain};
	
	\filldraw[fill=white, draw=black] (2.9,-2.6) rectangle (1.4,-2);
	\node[align=center] at (2.15,-2.3) {RF chain};
	
	% Wiring Rx RF chains to Digital combiner
	\draw (2.9,-1.1) -- (3.4,-1.1);
	\node[current direction={rotate = 0}] at (3.3,-1.1) {};
	
	\draw (2.9,-2.3) -- (3.4,-2.3);
	\node[current direction={rotate = 0}] at (3.3,-2.3) {};
	
	% Digital Combiner block	
	\draw (3.4,-0.8) rectangle (5.2,-2.6);
	\node[align=center] at (4.3,-1.7) {Digital\\Combiner\\ \\$\mathbf{W}_K$};
	
	% Wiring Digital Combiner to Decoder
	
	\draw (5.7,-1.1) -- (5.2,-1.1);
	\node[current direction={rotate = 0}] at (5.6,-1.1) {};
	
	%\node[align=center] at (5.45,0.7) {$\vdots$};
	\node[align=center] at (5.45,-1.7) {$N_A$};
	%\node[align=center] at (5.45,-0.5) {$\vdots$};
	
	\draw (5.7,-2.3) -- (5.2,-2.3);
	\node[current direction={rotate = 0}] at (5.6,-2.3) {};

	% Decoder Block
	\filldraw[fill=white, draw=black] (5.7,-0.8) rectangle (7.5,-2.6);
	\node[align=center] at (6.6,-1.7) {Decoder};

	% Decoder output
	\draw (8,-1.7) -- (7.5,-1.7);
	\node[current direction={rotate = 0}] at (7.9,-1.7) {};		
	\node[align=center] at (8.7,-1.7){$\hat{a}_K, \hat{s}_K$};	
	
\end{tikzpicture}